\documentclass[twocolumn,amsmath,floatfix, showpacs,aps,amssymb,superscriptaddress]{revtex4-1}
\usepackage{graphicx}
\usepackage{dcolumn}
\usepackage{setspace}
\usepackage{color}
\begin{document}

\title{Thermodynamics of a stochastic twin elevator}

\author{Niraj Kumar}
 \affiliation{Department of Chemistry and Biochemistry and BioCircuits
Institute,
University of California San Diego,\\
9500 Gilman Drive, La Jolla, CA
92093-0340, USA\\}
\author{Christian Van den Broeck}
\affiliation{Hasselt University, B-3590 Diepenbeek, Belgium}
\author{Massimiliano Esposito}
\affiliation{Center for Nonlinear Phenomena and Complex Systems,
Universit\'e Libre de Bruxelles,\\
CP 231, Campus Plaine, B-1050 Brussels, Belgium}
\author{Katja Lindenberg}
\affiliation{Department of Chemistry and Biochemistry and BioCircuits
Institute,
University of California San Diego,\\
9500 Gilman Drive, La Jolla, CA
92093-0340, USA\\}

\date{\today}

\begin{abstract}
We study the non-equilibrium thermodynamics of a single
particle with two available energy levels,
in contact with a classical (Maxwell-Boltzmann) or quantum (Bose-Einstein)
heat bath. The particle
can undergo transitions between the levels via thermal activation or deactivation.  The
energy levels are alternately raised at a given rate regardless of occupation by 
the particle, maintaining a fixed energy gap equal to $\varepsilon$ between them.
We explicitly calculate the work, heat and entropy production rates.
The efficiency in both the classical and the quantum case
goes to a limit between $100\%$ and 
$50\%$ that depends on the relative rates of particle transitions and level elevation.
In the classical problem we explicitly find the large deviation functions for heat, work, and
internal energy.
\end{abstract}


\maketitle{}
\section{Introduction}
\label{one}
Over the past decade, there has been growing interest in the stochastic
energetics of small systems. This interest is driven in part by the impressive
experimental and technological progress in bio- and nano-technology. At the same time,
the study of small scale systems has led to spectacular developments in nonequilibrium
statistical mechanics and thermodynamics. Brownian motors and refrigerators~\cite{bmr},
work and fluctuation theorems~\cite{massimilianoreview,wft},
and stochastic thermodynamics~\cite{st1,st2}
provide prominent examples of these developments. Among the issues of specific interest
are the thermodynamic properties of small scale stochastic systems and, in particular,
the efficiency of interconverting different forms of energy. For classical heat
engines, a certain degree of universality has been identified for the transformation
of heat into work. In particular, the efficiency $\eta$ at maximum power is found to
be half of Carnot efficiency in the regime of linear response~\cite{universality}.
This result has been illustrated by explicit calculations for several small
scale engines~\cite{sse,we1,we2}. However, many  artificial and most biological engines
operate in an isothermal environment. They transform one form of energy
(e.g. chemical or electrical) into another form (e.g. mechanical or optical).
Thermodynamics prescribes that the efficiency of this transformation is at most
$100\%$, a limit again reached for a reversible, hence zero-power transformation.
Concerning the  efficiency at maximum power, it appears that there is again universality
at the lowest order, i.e., in the regime of linear response $\eta=1/2$~\cite{jstatmech}. This result
is reminiscent of the so-called maximum power transfer theorem  from electrical
engineering~\cite{mpt}, enunciated by Moritz von Jacobi around 1840:
maximum power is achieved when the load resistance is equal to the source resistance,
with corresponding efficiency equal to $50\%$.  The issue of universality beyond linear
response is currently under debate~\cite{wtw}.

The standard way to apply external work in statistical mechanics is to systematically
move (modulate) energy levels or to modulate the potential energy. Unfortunately, the
analytic treatment of even the simplest
case, namely, modulating a single energy level, appears to be extremely difficult, see
for example~\cite{we1,we2,ritort,comment}. The main purpose of this paper is to introduce
an exactly solvable toy model which can be solved in full analytic detail, both in
a classical and a quantum setting.  We will call it the
stochastic twin elevator. We present explicit results for the rates of work, heat
and entropy production.
The efficiency of conversion of external work into internal
energy is found to vary between $100\%$ and $50\%$.
For the case of a classical bath we also derive the analytic expressions
for the large deviation functions~\cite{thearXivref} that characterize the statistics
of the accumulated stochastic work, heat
and internal energy and show that a heat fluctuation theorem is satisfied in the steady
state. Such an explicit calculation is the exception rather than the rule~\cite{cleuren}. 

The paper is organized as follows.  In Sec.~\ref{two} we introduce the
stochastic twin elevator model and present the evolution
equations and associated
rates of heat, work, internal energy, and entropy
production, as well as the results for the efficiency of the energy conversion process.
In Sec.~\ref{three} we concentrate on the classical bath and derive the steady
state fluctuation theorem for heat. We also explicitly calculate the large deviation
functions for the heat, the work, and the internal energy.
We conclude with a brief summary in Sec.~\ref{four}.

\section{Stochastic twin elevator model}
\label{two}

The model is defined as follows. A single particle,
in contact with a heat bath, can reside in one of two
available energy levels separated by a fixed energy gap equal to $\varepsilon \equiv \epsilon/k_BT$.
Here $k_B$ is the Boltzmann constant and $T$ is the temperature of the
bath.  Thus $\varepsilon$ is the energy in units of $k_BT$; all energies will be expressed
in these units.  The levels are alternately and instantaneously raised at random times at
a rate $k_c$, while maintaining the fixed energy gap $\varepsilon$ between them.
When the particle occupies the level that is raised, the external agent must perform
an amount of work equal to $2\varepsilon$ on the system. If the level that is raised is unoccupied,
the external work is zero. Note that this raising of the levels is a disturbance that drives
the system away from equilibrium. Due to its contact with the heat bath, the particle can
at any time make a thermal transition from the level it is occupying to the other level,
absorbing from the bath
(if the transition is uphill) or releasing to the bath (for downhill transitions)
an amount of heat equal to $\varepsilon$. In this way, work and heat can be monitored.
At the same time, the entropy produced in the process is known from stochastic thermodynamics,
see below for more details.
 
\begin{figure}
\vspace*{0.5cm}
\includegraphics[width=8cm]{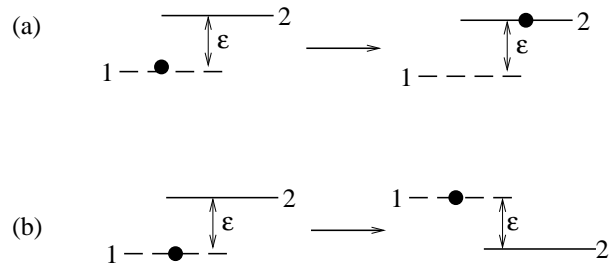}
\caption{Schematic representation of the two configurations of the twin elevator.
The dotted line represents level 1 while the solid one is level 2. 
(a) Hopping of particle in configuration 1 from the lower level to the higher level leading
to a transition from state (1,1) to (1,2). (b) Transition from configuration 1 to 2 due to lifting
of the lower energy level by $2\varepsilon$ thus changing the state from (1,1) to (2,1).
}
\label{fig:el_1}
\end{figure}

The technical simplicity of the stochastic twin elevator is due to the following
mapping onto a 4-state Markovian model. 
Let us arbitrarily call one of the elevator levels ``1" and the other ``2." 
The system can be in one of the four states $\{(1,1), (1,2), (2,1),(2,2)\}$, 
the first index indicating which state is the lower and the second indicating which state holds
the particle.  Thus, for  example, the state label $(1,2)$ means that
level 1 is below level 2, and that the particle occupies level 2 (the higher one in this case).
The system undergoes stochastic transitions between the four states.  For example, lifting
of the lower level in the state $(1,1)$ corresponds to a transition to the state $(2,1)$,
cf. Fig.~\ref{fig:el_1}. Similarly, if the particle in state $(1,1)$ makes a thermal jump
to the other energy level, the state changes to $(1,2)$. These transitions all
take place randomly in time, so that the probability distribution vector
${\mathbf P}$ with elements
\begin{equation}
{\mathbf P}=\begin{bmatrix} P(1,1)\\P(1,2)\\P(2,1)\\P(2,2)\end{bmatrix}
\end{equation}
evolves according to the master equation
\begin{equation}
\frac{d{\mathbf P}}{dt}={\mathbf M}{\mathbf P}.
\label{themasterequation}
\end{equation}
The analysis of the properties of the system is thus reduced to matrix algebra involving
a time-independent $4 \times 4$ matrix.

Whether the bath is classical or quantum mechanical,
the rates for thermal transitions of the particle obey the
detailed balance condition
\begin{equation}
\frac{W_{h\rightarrow l}}{W_{l\rightarrow h}}=e^{\varepsilon},
\label{detailedbalance}
\end{equation}
where $W$ stands for the transition rate, $h$ and $l$ stand for 
higher and lower energy levels respectively.

\subsection{Classical bath}
Consider first the case of a classical heat bath, and let $k$ be the transition rate from the higher
to the lower energy level in a given configuration.  The transition rate from the lower to the
higher is $ke^{-\varepsilon}$ (in general, $k$ could depend on temperature):
\begin{eqnarray}
W_{(1,1)\rightarrow (1,2)} &=& W_{(2,2)\rightarrow (2,1)} = ke^{-\varepsilon}, \nonumber\\
W_{(1,2)\rightarrow (1,1)}&=& W_{(2,1)\rightarrow (2,2)}=k.
\end{eqnarray}
As mentioned before, $k_c$ is the rate of lifting the lower  energy level.
We conclude that the transition matrix is given by the following expression:
\begin{widetext}
\begin{equation}
{\mathbf M}=\begin{bmatrix}
-(ke^{-\varepsilon}+k_c) & k & k_c & 0 \\
ke^{-\varepsilon} & -(k+k_c) & 0 & k_c \\
k_c & 0 & -(k_c+k) & ke^{-\varepsilon} \\
0 & k_c & k & -(ke^{-\varepsilon}+k_c)
\end{bmatrix}
=k\begin{bmatrix}
-(e^{-\varepsilon}+\xi) & 1 & \xi & 0 \\
e^{-\varepsilon} & -(1+\xi) & 0 & \xi \\
\xi & 0 & -(1+\xi) & e^{-\varepsilon} \\
0 & \xi & 1 & -(e^{-\varepsilon}+\xi)
\end{bmatrix}
\end{equation},
\end{widetext}
where we have introduced the dimensionless parameter $\xi=k_c/k$.  We will exhibit results in terms
of the three parameters $\varepsilon$, $\xi$, and $k$.
We will focus on  the steady state properties. The steady state probabilities
$P(1,1)$, $P(1,2)$, $P(2,1)$, $P(2,2)$, with $P(1,1)+P(1,2)+P(2,1)+P(2,2)=1$,
are found as the components of the right eigenvector of the matrix ${\mathbf M}$ associated
with the zero eigenvalue. Explicitly,
\begin{eqnarray}
P(1,1)=P(2,2)=\frac{e^{\varepsilon}(1+\xi)}{2\left[1+e^{\varepsilon}(1+2\xi)\right]},\nonumber\\
P(1,2)=P(2,1)=\frac{1+\xi e^{\varepsilon}}{2\left[1+e^{\varepsilon}(1+2\xi)\right]}.
\end{eqnarray}

In the absence of driving, $k_c=0$, the 
current between any two states vanishes. However, this is no longer so
when $k_c>0$. In this case the four steady state currents
between the states indicated by the subscripts are given by
\begin{eqnarray}
I&\equiv&
I_{(1,1)\rightarrow (1,2)}=
I_{(1,2)\rightarrow (2,2)}=
I_{(2,2)\rightarrow (2,1)}=
I_{(2,1)\rightarrow (1,1)} \nonumber\\
&=&P(1,1)ke^{-\varepsilon}-P(1,2)k \nonumber\\
&=&\frac{k\xi \left(1-e^{\varepsilon }\right)}{2\left[1+e^{\varepsilon}(1+2\xi)\right]}.
\end{eqnarray} 
Let us look more closely at the nonequilibrium thermodynamic properties. The heat flux to the 
system (recall that all energies are given in units of $k_B T$) is given by
 \begin{eqnarray}
  \dot Q=&&\varepsilon\left[W_{11\rightarrow 12}P(1,1) -W_{12\rightarrow 11}P(1,2)\right]\nonumber\\
 &&-\varepsilon\left[W_{21\rightarrow 22}P(2,1)- W_{22\rightarrow
21}P(2,2)\right]\nonumber\\
&=&k\varepsilon\left[e^{-\epsilon}\left(P(1,1)+P(2,2)\right) \right. \nonumber\\
&&-\left.\left(P(1,2)+P(2,1)\right)\right].
\end{eqnarray}
The rate of change of the work, that is, the power delivered to
the system, is 
 \begin{eqnarray}
 \dot{\mathcal{W}} &=& 2\varepsilon\left[ W_{11\rightarrow 21}P(1,1)+
W_{22\rightarrow 12}P(2,2)\right]\nonumber\\
&=&2\varepsilon k\xi\left[P(1,1)+P(2,2)\right],
 \end{eqnarray}
which reflects the fact that work in the amount of $2\varepsilon$ is performed on the system
when an occupied level is lifted.
Using the First Law of thermodynamics, the
increase in the internal energy of the system per unit time is just the sum of these two
contributions,
\begin{equation}
\dot U=\dot {\mathcal W}+\dot Q. 
\end{equation}
Finally, the rate of total entropy production associated with the master
equation is given by~\cite{st1}
\begin{equation}
 \dot S_i=\sum W_{\nu,j\rightarrow \nu^\prime,j^\prime}P(\nu,j)\log\frac{W_{\nu, j\rightarrow \nu^\prime,j^\prime}P(\nu,j)}{W_{\nu^\prime,j^\prime\rightarrow \nu,j}P(\nu^\prime,j^\prime)},
\end{equation}
where the summation is over all possible states.

\begin{figure}[h]
\includegraphics[width=8cm]{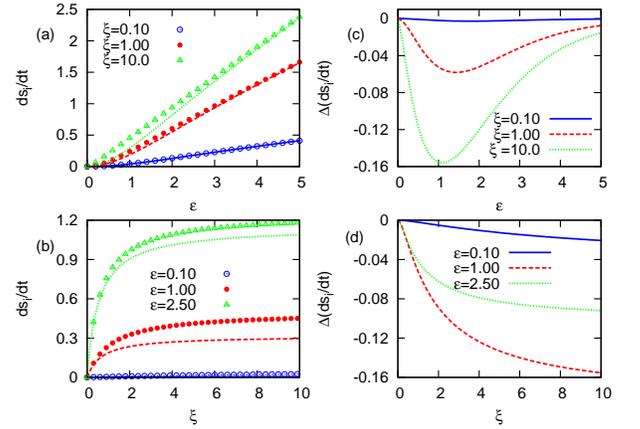}
\caption{(Color online) (a) Entropy production rate $\dot s_i$ as a function of $\varepsilon$
for different values of $\xi$. The lines are for a classical bath and the symbols for
a quantum bath.
(b) $\dot s_i$ versus $\xi$ for various values of $\varepsilon$. The lindes are for
a classical bath and the symbols for a quantum bath.
(c) Difference in entropy production rates between the classical and quantum bath cases
as a function of $\varepsilon$ for various values of $\xi$.
(d) Difference in entropy production rates between the classical and quantum bath cases
as a function of $\xi$ for various values of $\varepsilon$.
}
\label{fig:el_3}
\end{figure}

The thermodynamic quantities $\dot Q$, $\dot {\mathcal W}$, $\dot U$ and $\dot S_i$
in non-dimensional form $\dot q$, $\dot w$, $\dot u$ and $\dot s_i$ can be rewritten as
\begin{equation}{\label{eheat}}
  \dot q = \frac{\dot Q}{k\xi}=\frac{\varepsilon(1-e^\varepsilon)}{1+e^{\varepsilon}(1+2\xi)},
\end{equation}
\begin{equation}{\label{ework}}
 \dot w =\frac{\dot{\mathcal W}}{k\xi}=
  \frac{2 \varepsilon  e^{\varepsilon }(1+\xi)}{1+e^{\varepsilon}(1+2\xi)},
\end{equation}
\begin{equation}{\label{eu}}
  \dot u=\frac{\dot U}{k\xi}=\dot\omega+\dot q=\varepsilon,
\end{equation}
and
\begin{equation}
\dot s_i= -\xi \dot q.
\end{equation}
We have plotted the results for the entropy production rate as a function of
$\varepsilon$ and of $\xi$ in the two left panels of Fig.~\ref{fig:el_3}.
Note that equilibrium can be reached in two different ways,
namely, with $\varepsilon\rightarrow 0$ or with $\xi\rightarrow 0$. In these limits
$\dot s_i$  goes to zero. When the system is 
out of equilibrium, $\dot s_i$ increases when $\varepsilon$ or $\xi$ increase.
(The right panels will be discussed in the next subsection.)

\begin{figure}[h]
\includegraphics[width=8cm]{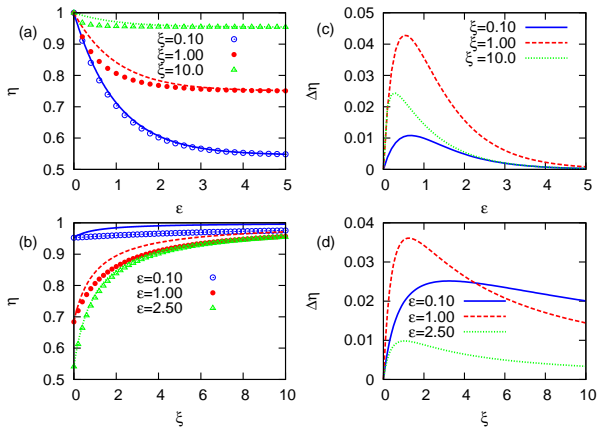}
\caption{(Color online) (a) Efficiency $\eta$ as a function of $\epsilon$ for
different values of $\xi$. The lines
are for a classical bath and the symbols for a quantum bath.
(b) $\eta$ versus $\xi$ for different values of $\varepsilon$. The lines are for
a classical heat bath and the symbols for a quantum bath.
(c) Difference of efficiencies between the classical and quantum bath cases as a function of
$\varepsilon$ for various values of $\xi$.
(d) Difference of efficiency between the classical and quantum bath cases as a function of $\xi$ for
various values of $\varepsilon$.
}
\label{fig:el_2}
\end{figure}

Finally, we turn to the efficiency of the system, which quantifies how efficiently the work
done on the system is utilized in increasing its internal energy and is given as 
\begin{equation}
\eta=\frac{\dot U}{\dot{\mathcal W}}=\frac{1+e^{\varepsilon}(1+2\xi)}{2e^\varepsilon(1+ \xi)}.
\end{equation}
In the two left panels of Fig.~\ref{fig:el_2} we have plotted the results for
efficiency as a function of $\varepsilon$ 
and of $\xi$ (the right panels will be discussed in the next subsection). We
observe that $\eta$  decreases with increasing
values of $\varepsilon$, cf. Fig.~\ref{fig:el_2}(a), and increases with increasing values of
$\xi$, cf. Fig.~\ref{fig:el_2}(b). We further note that:
\begin{list}{}
\item{(i)} For $\varepsilon\rightarrow 0$,
$\eta\rightarrow 1$. In this limit, all the four states have equal probability equal to $1/4$,
and so the efficiency $\eta=1$.
\item{(ii)} In the other extreme limit, i.e., 
$\varepsilon\rightarrow \infty$, $\eta\rightarrow 1/2$.
In this limit
\begin{equation}
\eta \to \frac{1+2\xi}{2(1+\xi)},
\end{equation}
which decreases from $1$ when $\xi \to \infty$ and goes to $1/2$ when $\xi \to 0$.
\item{(iii)} 
In general, there is a 
balance between the rate associated with configuration changes and that associated with particle
transitions. If the configuration changes very quickly compared to $k$, the efficiency of the system
increases. 
\end{list}

\subsection{Quantum bath}

The difference between the classical and the quantum versions of our toy model lies in the
nature of the bath.  In the former, the bath is described by Maxwell-Boltzmann statistics.  In the
latter, where, for example, the bath excitations might be phonons or photons,
the statistics are Bose-Einstein.  The rates at which the particle
makes a transition between the two levels in a given configuration involve the emission or
absorption of these excitations by the bath, and are now given by
\begin{eqnarray}
W_{(1,1)\rightarrow (1,2)}&=& W_{(2,2)\rightarrow (2,1)} =kn(\varepsilon)\nonumber\\
W_{(1,2)\rightarrow (1,1)} &=& W_{(2,1)\rightarrow (2,2)}
=k\left[1+n(\varepsilon)\right],
\end{eqnarray}
where, as before,
$\varepsilon$ is the energy difference between the levels in units of the thermal energy, 
$n(\varepsilon)=\left(e^{\varepsilon}-1\right)^{-1}$ is the Bose-Einstein distribution
function, and $k$ is a rate coefficient. Note that these transition elements obey the detailed 
balance condition Eq.~(\ref{detailedbalance}).
As in the classical case, the stochastic evolution of the system is described by the
master equation Eq.~(\ref{themasterequation}), but now with the transition matrix
\begin{widetext}
\begin{equation}
{\bf M}=k\begin{bmatrix}
-[n(\varepsilon)+\xi]	 & [n(\varepsilon)+1] &  \xi & 0\\
 n(\varepsilon ) & -[1+n(\varepsilon) +\xi] & 0 & \xi \\
\xi & 0 & -[1+n(\varepsilon)+\xi] & n(\varepsilon )\\
0 & \xi &[n(\varepsilon)+1] & -[n(\varepsilon)+\xi].
\end{bmatrix}.
\end{equation}
\end{widetext}
The master equation leads to the steady state solution for the probabilities 
\begin{eqnarray}
 P(1,1)=P(2,2)&=&\frac{e^{\varepsilon }(1+\xi)-\xi }
{2(2 e^\varepsilon\xi +e^\varepsilon -2 \xi +1)},
\nonumber\\
 P(1,2)=P(2,1)&=&\frac{e^{\varepsilon } \xi +1-\xi}
{2(2 e^\varepsilon\xi +e^\varepsilon -2 \xi +1)}.
\end{eqnarray}

Following our earlier rules for the classical case, we can write the rates of heat,
work, and internal energy influx into the system.
In adimensional form we find
 \begin{eqnarray}
\dot q&=&\frac{\varepsilon(1-e^{\varepsilon})}
{2 e^\varepsilon\xi +e^\varepsilon -2 \xi +1}, \nonumber\\
 \dot w&=&\frac{
2 \varepsilon \left(e^\varepsilon \xi +e^\varepsilon -\xi \right)}
{2 e^\varepsilon\xi +e^\varepsilon -2 \xi +1}, \nonumber\\
 \dot u&=&\dot \omega+\dot q=\varepsilon\nonumber\\
\dot{s}_i &=& -\xi \dot q.
 \end{eqnarray}
In the left panels of Fig.~\ref{fig:el_3} we show the entropy production rate as a function of
$\varepsilon$ (upper panel, symbols) and of $\xi$ (lower panel, symbols).
In the right hand panels of Fig.~\ref{fig:el_3} we
show the difference between the classical and quantum entropy production rates, in panel (c) as a
function of $\varepsilon$ for different values of $\xi$, and in panel (d) as a function of $\xi$ for
different values of $\varepsilon$.  The classical and quantum entropy production rates are equal 
in the limits $\varepsilon \to 0$ and $\varepsilon \to \infty$, but between these two limits the
quantum entropy production rate is everywhere greater than in the classical case.  As a function of
$\xi$ for fixed $\varepsilon$, the two again become equal as $\xi \to 0$. As $\xi \to \infty$
the difference goes to the limit $-\varepsilon e^{-\varepsilon}/2$.

Finally, we calculate the efficiency of the system: 
\begin{equation}
\eta=\frac{\dot U}{\dot{\mathcal W}}=\frac
{2 e^\varepsilon\xi +e^\varepsilon -2 \xi +1}
{2 \left(e^\varepsilon \xi +e^\varepsilon -\xi \right)}.
\end{equation}
The efficiency as a function of $\varepsilon$ and of $\xi$ are shown by the symbols in the left hand
panels of Fig.~\ref{fig:el_2}.
We show the difference between the classical and quantum efficiencies (classical minus quantum) in
the right hand panels of Fig.~\ref{fig:el_2} for a number of parameter values. 
In the limits $\varepsilon \to 0$ and $\varepsilon \to \infty$
the difference between the two goes to zero, as seen in panel (c), as it should.
In the limit $\varepsilon\rightarrow\infty$ we find
\begin{equation}
\eta=\frac{2 \xi +1}{2 \xi +2},
\end{equation}
again exactly as in the classical case. 
The approach of the classical and quantum results to one another with increasing $\varepsilon$ is
seen in panel (c) of the figure. 
For all values of $\varepsilon$ between these limits the efficiency is higher in the classical case.
As in the classical case, there is in the quantum case a 
balance between the rate associated with configuration changes and that associated with particle
transitions; if the configuration changes very quickly compared to $k$, the efficiency of the system
increases.  In panel (d) we show the difference between classical and quantum efficiencies as a
function of $\xi$ for various values of $\varepsilon$.  Again, the efficiencies are equal when
$\xi=0$, as they should be.  As $\xi$ increases, the difference goes through a maximum (it is always
higher in the classical case) and goes to zero again as $\xi\to \infty$ and both efficiencies go to
unity. 

\section{Fluctuation theorem and large deviation functions for classical bath}
\label{three}
We now turn to the classical case to explicitly calculate a number of other thermodynamic
properties for the twin elevator system, in particular, the large deviation function and
the steady state fluctuation theorem for heat, as well as the large deviation functions for work
and internal energy.

\subsection{Fluctuation theorem for $Q$}

Let $Q$ be the accumulated heat transferred to the reservoir up to time $t$. Since the 
transition of the particle between the levels is stochastic, the total accumulated heat $Q$
is also stochastic. Let $P(i,Q,t)$ be the probability that the system is in state $i$ at
time $t$ and the heat transferred to the reservoir is $Q$. The evolution of $P(i,Q,t)$ follows
from the equation
\begin{eqnarray}
 P(i,Q,t+dt)&=&\sum_{j}W_{j\rightarrow i}dt P(j,Q-\Delta Q_{j\rightarrow i},t)\nonumber\\
 &+&\left[1-\sum_{j}W_{i\rightarrow j}dt\right]P(i,Q,t).
\label{eq1}
\end{eqnarray}
Here $\Delta Q_{j\rightarrow i}$ is the amount of heat transferred to the reservoir as a result of 
a transition of the particle between the levels in
state $j$, the resultant state being state $i$,
with rate $W_{j\rightarrow i}$. As always,
the quantities $\Delta Q_{j\rightarrow i}$ and $Q$ are in units
of $k_BT$. The differential form of the evolution equation follows from the master
equation~(\ref{themasterequation}),
\begin{eqnarray}
\frac{\partial P(i,Q,t)}{\partial t}&=&\sum_{j}W_{j\rightarrow i}P(j,Q-\Delta Q_{j\rightarrow i},t)\nonumber\\
         &-&\sum_{j}W_{i\rightarrow j} P(i,Q,t).
\label{eq1a}
\end{eqnarray}
We solve the above equation using the characteristic function (the subscript $q$ labels this as the
characteristic function for the heat),
\begin{equation}{\label{eq2}}
\rho_q(i,\lambda,t)=\int_{-\infty}^{\infty} dQ e^{-\lambda Q}P(i,Q,t),
\end{equation}
whose evolution equation is obtained directly from Eq.~(\ref{eq1a}),
\begin{eqnarray}{\label{eq3}}
\frac{\partial \rho_q(i,\lambda,t)}{\partial t}&=&\sum_{j}W_{j\rightarrow i}
e^{-\lambda \Delta Q_{j\rightarrow i}}\rho_q(j,\lambda,t)\nonumber\\
         &-&\sum_{j}W_{i\rightarrow j} \rho_q(i,\lambda,t).
\end{eqnarray}
In matrix notation, it can be written as
\begin{equation}{\label{eq4}}
\frac{\partial {\mathbf \rho}_q}{\partial t}={\mathbf M}_q(\lambda){\mathbf \rho}_q,
\end{equation}
where  ${\mathbf M}_q(\lambda)$ is
\begin{widetext}
\begin{equation}
{{\mathbf M}_q(\lambda)}=k\left
[\begin{array}{cccc}
 -(e^{-\varepsilon}+\xi)& e^{-\lambda \varepsilon } & \xi & 0 \\
 e^{(\lambda-1)\varepsilon }  & -( 1+\xi) & 0 & \xi \\
 \xi & 0 & -( 1+\xi) & e^{(\lambda-1)\varepsilon } \\
 0 & \xi & e^{-\lambda \varepsilon }  & -(e^{-\varepsilon}+\xi)
\end{array}
\right],
\end{equation}
\end{widetext}
and 
$\rho_q$ is a column matrix with four components.
In writing the above matrix we make use of the fact that during an upward transition
of the particle in a given configuration the reservoir loses heat
to the system and so $q$ is $-\varepsilon$, and when the particle makes a downward transition
heat flows to the reservoir, i.e., $q=+\varepsilon$. We also note that the matrix 
$\mathbf M_\lambda$ and its adjoint $\mathbf M_\lambda^\dagger$ satisfy the symmetry
relation~\cite{lebowitz}
\begin{equation}
{\mathbf M}_q(\lambda) ={\mathbf M}_q^\dagger(1-\lambda).
\label{symmetryrelation}
\end{equation}

The matrix ${\mathbf M}_q(\lambda)$ has four eigenvalues, and the general solution
of Eq.~(\ref{eq4}) is written as a linear combination of the four independent
associated eigenvectors. However, the large $t$ behavior of ${\mathbf \rho}_q$ is dominated
by its largest eigenvalue, i.e.,
\begin{equation}{\label{eq6}}
 \langle e^{-\lambda Q}\rangle=\int_{-\infty}^{\infty}dQ e^{-\lambda Q}P(Q,t)
\sim e^{tf_q(\lambda)},
\end{equation}
where $f_q(\lambda)$ is the largest eigenvalue of the matrix $\mathbf M_q(\lambda)$,
which we find to be
\begin{widetext}
\begin{equation}{\label{eq7}}
f_q(\lambda)=\frac{ke^{-\varepsilon}}{2} \left(
-(2\xi e^\varepsilon + e^\varepsilon +1) +\sqrt{
4\xi e^\varepsilon\left( \xi e^\varepsilon + e^{\lambda \varepsilon} +
e^{(1-\lambda)\varepsilon}\right) + (1+e^\varepsilon)^2 }\right).
\end{equation}
\end{widetext}
We observe that the eigenvalue $f_q(\lambda)$ obeys the symmetry relation
\begin{equation}{\label{eq9}}
f_q(\lambda)=f_q(1-\lambda),
\end{equation}
which is a direct consequence of the symmetry relation~(\ref{symmetryrelation})
for the matrix  ${\mathbf M}_q(\lambda)$.
The above symmetry relation for the maximum eigenvalue reflects the steady state
fluctuation theorem~\cite{massimilianoreview}, i.e.,
\begin{equation}{\label{eq10}}
\frac{P(Q)}{P(-Q)}\sim e^Q.
\end{equation}

It is easy to verify that the average heat per unit time released to the reservoir is given
in terms of the first derivative of the maximum eigenvalue evaluated at $\lambda=0$, i.e.,
\begin{equation}{\label{eq11}}
\langle \dot Q \rangle =-\frac{df_q(\lambda)}{d\lambda}|_{\lambda=0}=
\frac{k\xi\varepsilon \left(e^\varepsilon-1\right)}{1+ e^{\varepsilon }(1+2 \xi)}.
\end{equation}
We note that the magnitude of $\langle \dot Q \rangle $ is the same as that given
in Eq.~(\ref{eheat}) (where the average is understood).

We next turn to the explicit evaluation of the large deviation properties. For this we
need to find the 
probability $P(Q,t)$ for long times. The heat $Q$ is expected to grow linearly in time.
We thus introduce the variable $\phi=Q/t = \dot Q$, which is the heat flux. The flux $\phi$ can
be positive as well as 
negative due to gain or loss of heat. According to large deviation theory~\cite{thearXivref},
the probability
$P(Q,t)$ at long times can be written as
\begin{equation}{\label{eq12}}
 P(Q,t)\sim e^{-tg_q(\phi)},
\end{equation}
where $g_q(\phi)$ is the large deviation function. To find the relation between
$f_q(\lambda)$ and $g_q(\phi)$,
we implement the change of variables from $Q$ to $\phi$ in the integrand of Eq.~(\ref{eq6}),
which immediately allows us to identify $f_q(\lambda)$ with the extremum of
$-(g_q(\phi)+\lambda\phi)$ with respect to $\phi$. 
That is, invoking large deviation theory we see that the
maximum eigenvalue $f_q(\lambda)$ and the large deviation function
$g_q(\phi)$ are related by a Legendre transformation, leading to:
\begin{equation}{\label{eq14}}
g_q(\phi)=-(f(\lambda_\phi)+\phi\lambda_\phi),
\end{equation}
 where $\lambda_\phi$ is the solution of 
\begin{equation}{\label{eq15}}
f_q^\prime(\lambda_\phi)+\phi=0,
\end{equation}
the prime denoting differentiation with respect to $\lambda$. 
We find that the large deviation function $g_q(\phi)$ is given by (see the Appendix)
\begin{eqnarray}{\label{eq16}}
 g_q^{\pm}(\phi)&=&\frac{ke^{-\varepsilon}}{2}\left[-(2\xi e^\varepsilon + e^\varepsilon +1) 
\right.\nonumber\\
&&\left.-\sqrt{4\xi^2 e^{2\varepsilon} + (1+e^\varepsilon)^2 +4\xi e^{\varepsilon}
\left(\gamma^{\mp}(\phi)+\frac{e^{\varepsilon}}{\gamma^{\mp}(\phi)}\right)}\right]\nonumber\\
&&-\frac{\phi}{\varepsilon}\ln \gamma^{\mp}(\phi),
\end{eqnarray}
where we have introduced the distinct notation $g_q(\phi)\equiv g_q^{+}(\phi)$ when $\phi>0$
and $g_q(\phi)\equiv g_q^{-}(\phi)$ when $\phi<0$, and 
\begin{equation}{\label{eq17}}
 \gamma^{\mp}(\phi)=\frac{1}{2}\left(p(\phi)\mp\sqrt{p^2(\phi)-4e^{\varepsilon}}\right),
\end{equation}
with 
\begin{widetext}
\begin{equation}{\label{eq18}}
 p(\phi)=\frac{\sqrt{4 k^4 \xi^2\varepsilon^4 e^{-\varepsilon } +4k^2 \varepsilon ^2 \phi ^2\xi^2
+ k^2\varepsilon^2\phi^2 e^{-2\varepsilon} (1+e^\varepsilon)^2
+4 \phi ^4}+2 \phi ^2}{k^2 \varepsilon ^2 \xi e^{-\varepsilon}}.
\end{equation}
\end{widetext}

\begin{figure}
\includegraphics[width=8cm]{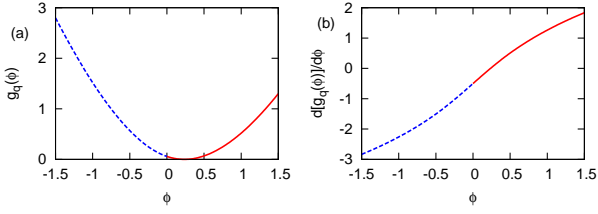}
\caption{(Color online) Large deviation function $g_q(\phi)$ and its derivative  
         as a function of $\phi$ for $\varepsilon=1$, $\xi=2$, $k=1$. The solid
line is for $g_q^{+}(\phi)$
         while the dashed line is for $g_q^{-}(\phi)$. The value of $\phi$ where $g_q(\phi)=0$
is 0.23552, which is the same as $\langle \dot Q\rangle$ obtained from Eq.~(\ref{eq11}).}
\label{fig:large-devQ}
\end{figure}

In Fig.~\ref{fig:large-devQ} we show the large deviation function and its derivative as a
function of $\phi$ for particular parameter choices. We note that the functions $g_q^{+}(\phi)$
and $g_q^{-}(\phi)$ match smoothly at $\phi=0$ as their values are the same at that point,
\begin{widetext}
\begin{equation}{\label{eq19}}
g_q^+(0)=g_q^{-}(0) =-\frac{1}{2} k e^{-\varepsilon}\left(
2\xi e^\varepsilon+e^\varepsilon+1 -
\sqrt{8 \xi e^{3\varepsilon/2}+4\xi^2e^{2\varepsilon}+(1+e^\varepsilon)^2}
\right),
\end{equation}
\end{widetext}
as are their first derivatives,
\begin{equation}{\label{eq20}}
\frac{dg^+(0)}{d\phi}=\frac{dg^-(0)}{d\phi}=-\frac{1}{2}.
\end{equation}
We also note that $g_q(\phi)$ has a single minimum at $\phi=\langle \dot Q \rangle$, as given
by  Eq.~(\ref{eq11}), and
at that point $g_q(\phi)=0$. Although the large deviation function is
a complicated nonlinear function of $\phi$, 
the difference between $g_q^+(\phi)$ and $g_q^-(\phi)$
turns out to be a simple linear function. That is, for $\phi>0$,
\begin{equation}{\label{eq}}
g_q(-\phi)-g_q(\phi)=g_q^{-}(-\phi)-g_q^{+}(\phi)=\phi,
\end{equation}
which leads to the fluctuation theorem for heat as written in Eq.~(\ref{eq10}).

\subsection{Large deviation function for  $\mathcal{W}$}

We next turn to large deviation function for the work $\mathcal{W}$ done on the system.
Work is done on the system whenever the filled lower energy level is lifted by energy
$2\varepsilon$.  The process of lifting the
level at rate $k_c$ is a random Markov process. As time increases, the work
$\mathcal{W}$ also increases. Let
$P(i,\mathcal{W},t)$ be the probability that the system is in
state $i$ at time $t$  with total accumulated
 work $\mathcal{W}$. In analogy with the heat in Eq.~(\ref{eq1a}), the Markovian
nature of the evolution allows us to write
\begin{eqnarray}
\frac{\partial P(i,\mathcal{W},t)}{\partial t}&=&\sum_{j}W_{j\rightarrow i}P(j,\mathcal{W}-\Delta\mathcal{W}_{j\rightarrow i},t)\nonumber\\
         &&-\sum_{j}W_{i\rightarrow j} P(i,\mathcal{W},t).
\end{eqnarray}
Here $\Delta\mathcal{W}_{j\rightarrow i}=2\varepsilon$ is the work done on the system when its
state change from $j$ to $i$ at rate $W_{j\rightarrow i}=k_c$. The characteristic function 
is given by
\begin{equation}
\rho_w(i,\lambda,t)=\int_{-\infty}^{\infty} d\mathcal{W} e^{-\lambda \mathcal{W}}P(i,\mathcal{W},t),
\end{equation}
with subscript $w$ standing for work. The evolution equation for $\rho_w(i,\lambda,t)$ is
\begin{eqnarray}
\frac{\partial \rho_w(i,\mathcal{W},t)}{\partial t}&=&\sum_{j}W_{j\rightarrow i}e^{-\lambda \Delta\mathcal{W}_{j\rightarrow i}}\rho_w(j,\lambda,t)\nonumber\\
         &&-\sum_{j}W_{i\rightarrow j} \rho_w(i,\lambda,t),
\end{eqnarray}
or, equivalently,
\begin{equation}
\frac{\partial \rho_w}{\partial t}={\mathbf M}_w(\lambda)\rho_w,
\end{equation}
where
\begin{widetext}
\begin{equation}{\label{ewmatrix}}
{\mathbf{M}}_w({\lambda})=k\left[
\begin{array}{cccc}
 -(e^{-\varepsilon}+\xi)&  1 & \xi & 0 \\
 e^{-\varepsilon} & -( 1+\xi) & 0 & e^{-2 \lambda \varepsilon } \xi \\
 e^{-2 \lambda \varepsilon } \xi & 0 & -( 1+\xi) & e^{-\varepsilon} \\
 0 & \xi &  1 & -(e^{-\varepsilon}+\xi)
\end{array}
\right].
\end{equation}
\end{widetext}
Again, the large $t$ behavior of $\rho_w$ is dominated by the maximum eigenvalue,
i.e.,
\begin{equation}
\langle e^{-\lambda \mathcal{W}}\rangle \sim e^{t{f_w(\lambda)}},
\end{equation}
where $f_w(\lambda)$ is the largest eigenvalue of the matrix
$\mathbf{M}_w(\lambda)$, 
\begin{widetext}
\begin{equation}
f_w(\lambda)=\frac{1}{2} ke^{-\varepsilon}
\left[
-(2\xi e^\varepsilon + e^\varepsilon +1)
e^{-\lambda\varepsilon} \sqrt{(4 \xi
e^{\varepsilon} +1) e^{2 \lambda \varepsilon }+e^{2
    (\lambda +1)\varepsilon}+2 e^{ (2\lambda +1)\varepsilon }+4 e^{2\varepsilon } \xi +4 \xi
   ^2 e^{2\varepsilon}}
\right].
\end{equation}
\end{widetext}
We can verify that the average work done per unit time,
\begin{equation}{\label{ewav}}
\langle \dot {\mathcal{W}}\rangle=-\frac{df_w(0)}{d\lambda}=\frac{2 k\varepsilon e^{\varepsilon
}\xi \left( 1+\xi \right)}{1+e^{\varepsilon }(1+2 \xi)},
\end{equation}
is the same as that obtained in Eq.~(\ref{ework}).

Next we use the large deviation theory to shed more light on the probability
distribution $P(\mathcal{W},t)$, which allows one to write,
\begin{equation}
P(\mathcal{W},t)\sim e^{-tg_w(\phi)},
\end{equation}
where $g_w(\phi)$ is the large deviation function
and $\phi$ is the flux of work ($\phi=\mathcal{W}/t = \dot {\mathcal{W}}$).
Here we note that (contrary to the case of heat) $\phi$ is always positive as the
system always receives work so that $\mathcal{W}$ increases monotonically with time.
The function $g_w(\phi)$ is again related to $f_w(\lambda)$ by its Legendre transform.
Following the same steps as for the heat, we now find the large deviation function
$g_w(\phi)$ to be given by
\begin{widetext}
\begin{equation}
 g_w(\phi)=\frac{1}{2\zeta_1}\left[k e^{-\varepsilon}[(2\xi e^\varepsilon + e^\varepsilon +1)
\zeta_1-\zeta_2]-\frac{\phi}{\varepsilon}\zeta_1\ln2\zeta_1^2\right],
\end{equation}
where
\begin{equation}
 \zeta_1=e^\varepsilon\sqrt{\frac{\xi (1+\xi)\left[-\phi+\sqrt{k^2\epsilon^2e^{-2\epsilon}
\left(4\xi e^\varepsilon + (1+e^\varepsilon)^2
\right)
+\phi^2}\right]}{
  \left(
4\xi e^\varepsilon + (1+e^\varepsilon)^2
\right)
\phi}},
\end{equation}
and 
\begin{equation}
 \zeta_2=e^\varepsilon \sqrt{\frac{\xi (1+\xi)\left[\phi+\sqrt{k^2\epsilon^2
e^{-2\epsilon}
\left(
4\xi e^\varepsilon +(1+e^\varepsilon)^2
\right)
+\phi^2}\right]}{\phi}}.
\end{equation}
\end{widetext}

\begin{figure}
\includegraphics[width=8cm]{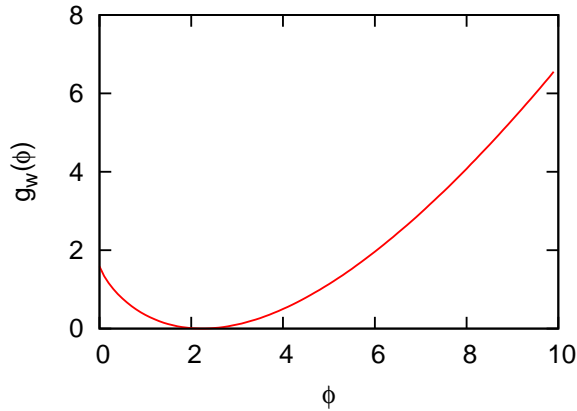}
\caption{(Color online) Large deviation function for work. Parameters are: $\epsilon=1, \xi=2, k=1$.
The value of $\phi$ where $g_w(\phi)=0$ is unique and its value is 0.23552, which is
the same as that obtained from Eq.~(\ref{ewav}).}
\label{fig:large-dev-w}
\end{figure}
In Fig.~\ref{fig:large-dev-w} we have plotted the large deviation function for work for
a particular set of parameters. We see that it
has a unique minimum and $g_w(\phi)=0$ at $\phi=\langle \mathcal{W}/t\rangle$.

\subsection{Large deviation function for the internal energy}

\begin{figure}
\includegraphics[width=8cm]{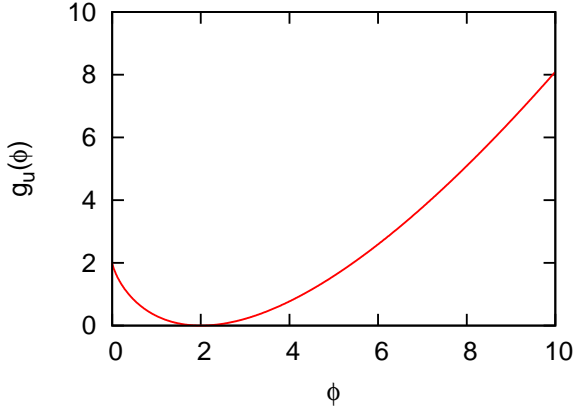}
\caption{(Color online) Large deviation function for the internal energy. Parameters are: $\epsilon=1,
\xi=2, k=1$.
The value of $\phi$ where $g_u(\phi)=0$ is unique and its value is 2, which is
the same as that obtained from Eq.~(\ref{eint}).}
\label{fig:large-dev-u}
\end{figure}

Finally, we turn to the large deviation function for the internal energy. Following the
procedure in the previous sections, we can write the evolution of the characteristic function,
\begin{equation}
\rho_u(i,\lambda,t)=\int_{-\infty}^{\infty} dU e^{-\lambda U}P(i,U,t),
\end{equation}
(subscript $u$ for internal energy), as
 \begin{eqnarray}
 \frac{\partial \rho_u(i,U,t)} {\partial t} &=&\sum_{j}W_{j\rightarrow i}e^{-\lambda \Delta U_{j\rightarrow i}}\rho_u(j,\lambda,t)\nonumber\\
          &&-\sum_{j} W_{i\rightarrow j} \rho_u(i,\lambda,t).
 \end{eqnarray}
Here $\Delta U$ is the change in internal energy when the system undergoes a transition 
from one state to the other either due to a jump of the particle between the levels
or by lifting the filled lower level.
The above evolution equation can be rewritten as the matrix equation
\begin{equation}
\frac{\partial \rho_u}{\partial t}=\mathbf{M}_u(\lambda)\rho_u,
\end{equation}
where
\begin{widetext}
\begin{equation}
\mathbf M_{u}(\lambda)=k\left[
\begin{array}{cccc}
 -(e^{-\varepsilon}+\xi) & e^{\lambda \varepsilon } & \xi & 0 \\
 e^{- (\lambda +1)\varepsilon }  & -(1+\xi)& 0 & e^{-2 \lambda \varepsilon }\xi \\
 e^{-2 \lambda \varepsilon } \xi & 0 & -(1+\xi)& e^{-(\lambda +1)\varepsilon }  \\
 0 & \xi & e^{\lambda \varepsilon }  & -( e^{-\varepsilon}+\xi)
\end{array}
\right].
\end{equation}
\end{widetext}
In writing the above matrix, we have made use of the fact that during an upward transition
of the particle,
there is a gain of internal energy ($\Delta U=\varepsilon$) and, during a downward transition 
there is a loss
($\Delta U=-\varepsilon$). There are four such transitions where the system internal
energy increases or decreases as a result of particle transitions between the levels.
Similarly, when the lower level containing the particle 
is lifted, the internal energy of the system increases, $\Delta U=2\varepsilon$. There
are two such transitions that lead to an increase in the internal energy of
the system by lifting. We are interested
in the long time behavior, and so we can write
\begin{equation}
\langle e^{-\lambda U}\rangle \sim e^{t{f_u(\lambda)}},
\end{equation}
where $f_u(\lambda)$ is the largest eigenvalue of matrix $\mathbf M_u(\lambda)$ and is given as
\begin{equation}
f_u(\lambda)=k \xi  \left(e^{-\lambda \varepsilon }-1\right).
\end{equation}
We verify that 
\begin{equation}{\label{eint}}
\langle \dot U \rangle=-\frac{df_u(0)}{d\lambda}=k\xi\varepsilon,
\end{equation}
in agreement with Eq.~(\ref{eu}). Next, we use large deviation theory 
to write for the probability distribution for the internal energy
\begin{equation}
P(U,t)\sim e^{-tg_u(\phi)},
\end{equation}
where $g_u(\phi)$ is the large deviation function for the internal energy and $\phi=U/t = \dot U$.
The large deviation function and maximum eigenvalue are again related by a Legendre
transform, from which we obtain
\begin{equation}
g_u(\phi)=\frac{1}{\varepsilon}
\left(\phi  \log \left(\frac{\phi }{k \varepsilon  \xi }\right)+k \varepsilon  \xi -\phi \right).
\end{equation}
In Fig.~\ref{fig:large-dev-u} we show the large deviation function for the internal energy for
a particular set of parameter values. 
Again, we note that it has a unique minimum at $\phi=\langle \dot U \rangle $ where $g_u(\phi)=0$.

\section{Conclusion}
\label{four}

We have presented the stochastic thermodynamics of a single 
particle that can reside on one of two energy levels, in contact with a classical
or a quantum heat bath. The energy levels have a fixed energy separation and the system is
driven out of equilibrium by alternately and stochastically lifting one of the two energy levels.
The particle can make upward or downward transitions between the energy levels mediated
by the heat bath. At a given bath temperature, three parameters determine the behavior
of the system: the energy separation
between the levels, the transition rates for the particle to move between levels, and the rate at
which the levels are stochastically raised. 
The interest of this toy model lies in the fact that we can obtain explicit analytic expressions
not only for average quantities such as work and heat flux, rate of entropy production and
efficiency, but also of stochastic trajectory-dependent quantities including the steady state
fluctuation theorem for the heat and the
large deviation properties of work, heat and internal energy,
both at the level of their characteristic function as well as in
terms of the variables themselves.

\appendix
\section{Large deviation function}
\label{appendixQ}
To find the large deviation function, using Eq.~(\ref{eq15}) we must solve the equation
\begin{equation}{\label{ea1}}
 \left(\frac{df_q(\lambda)}{d\lambda}\right)^2=\phi^2.
\end{equation}
Differentiating $f_q(\lambda)$ with respect to  $\lambda$ gives
\begin{equation}{\label{ea2}}
\frac{df_q(\lambda)}{d\lambda}=\frac{\xi\varepsilon k\left(e^{\lambda\varepsilon} -
e^{-(\lambda-1)\varepsilon}\right)}
{\sqrt{4\xi^2 e^{2\varepsilon}
+ (e^\varepsilon+1)^2
+4\xi\left(e^{\lambda\varepsilon}   
+ e^{(1-\lambda)\varepsilon}\right)}}.
\end{equation}
Using Eq.~(\ref{ea2}) in Eq.~(\ref{ea1}), we get the following quadratic equation in $p$:
\begin{equation}{\label{ea4}}
\xi^2\varepsilon^2 k^2p^2-4\xi e^\varepsilon\phi^2p-\left(4\xi^2\varepsilon^2k^2e^\varepsilon+\phi^2\psi_2\right)=0,
\end{equation}
where
\begin{equation}{\label{ea5}}
p= \gamma + \frac{e^\varepsilon}{\gamma}, \qquad \gamma \equiv e^{\lambda \varepsilon}.
\end{equation}
$p$ is always positive since it is a sum of exponentials. We thus consider only the
positive solution of Eq.~(\ref{ea4}), which is given by Eq.~(\ref{eq18}). We note 
that $p$ is an even function of $\phi$, i.e., $p(\phi)=p(-\phi)$, and is an increasing 
function of $\phi$. The limiting values of $\phi$ are,
\begin{eqnarray}{\label{ea6}}
p(\phi)&=&2e^{\epsilon/2}  ~~\text{for}~~ \phi\rightarrow 0,\nonumber\\
       &=&\infty ~~\text{for}~~ \phi\rightarrow\pm\infty.
\end{eqnarray}

Using Eq.~(\ref{ea5}), we get a quadratic equation in $\gamma(\phi)$ with two 
roots $\gamma^{-}(\phi)$ and $\gamma^{+}(\phi)$ as given by Eq.~(\ref{eq17}). The
valid solution for $\gamma(\phi)$ must satisfy Eq.~(\ref{eq15}). That is,
for $\phi>0$, $df_q(\lambda)/d\lambda$ must be negative, which from Eq.~(\ref{ea2})
requires that $e^{(2\lambda-1)\varepsilon} <1$. Using the properties and limiting values of
$p(\phi)$ as given in Eq.~(\ref{ea6}), it turns out that this inequality 
is satisfied only if $\gamma(\phi)=\gamma^{-}(\phi)$. Similarly,
for $\phi<0$, $\gamma(\phi)=\gamma^{+}(\phi)$. Finally, using these values of $\gamma(\phi)$ in 
Eq.~(\ref{eq14}), we get the desired large deviation function $g_q(\phi)$ as
given in Eq.~(\ref{eq16}).


\begin{thebibliography}{}
\bibitem{comment}
To illustrate the level of complication, we note that the master equation for a modulated
single level system with two states, empty or occupied, involves a time-dependent
two by two matrix. For periodic modulation, the calculation of even the simplest
properties (time-averaged quantities) is mathematically analogous to the discussion of
parametric oscillators.
\end{thebibliography}
\end{document}